\documentstyle[prl,aps,epsf,floats]{revtex}

\begin{document}
\draft

\twocolumn[\hsize\textwidth\columnwidth\hsize\csname @twocolumnfalse\endcsname

\title{
Comment on `Origin of combination frequencies in quantum magnetic oscillations 
of two-dimensional multiband metals' 
by T. Champel [Phys. Rev. B {\bf 65}, 153403 (2002)]
}
\author{
A.S. Alexandrov$^{1,2}$ and  A.M.  Bratkovsky$^1$
}
\address{$^1$Hewlett-Packard Laboratories, 1501 Page Mill 
Road, 1L, Palo Alto, California 94304\\
$^2$Department of Physics, Loughborough University,
Loughborough LE11 3TU, United Kingdom}
\maketitle

\begin{abstract}
We analyze the applicability of our analytical theory of  combination
harmonics in canonical low-dimensional multi-band Fermi liquids,
which was recently criticized by Champel (Phys. Rev. B {\bf 65}, 153403 (2002)).
It is shown that his claim that our  analytical theory does not apply at low
temperatures and in clean samples, is incorrect. 
We demonstrate that
the analytical theory of combination harmonics is in excellent agreement
with the  exact numerical results even at zero temperature and for
clean systems,
which are the most challenging for an analytical description.
\end{abstract}

\vskip2pc]

\narrowtext

The combination frequencies in the magnetization of the two-dimensional
multiband metals with constant net electron density have been predicted,
given simple physical explanation, and described numerically in Ref.~\cite
{alebra}. They were further numerically studied in Refs.\cite{alebra2,nak}
and observed experimentally by Shepherd {\it et al. }\cite{she}{\it \ } and
by other groups \cite{other}. We have also proposed an analytical theory 
\cite{alebra3} of the combination Fourier components in the framework of the
semiclassical Lifshitz-Kosevich approach \cite{kos}. More recently the
theory has been extended by taking into account the Dingle, spin and angle
(Yamaji) reduction factors, and the nonquantized ``background'' density of
states \cite{alebra4}. The purpose of the present paper is to analyze our
analytical theory in view of critical remarks in Ref.\cite{cham}, who
claimed that ``{\em the chemical potential oscillations appearing in the
arguments of the Fourier components were not taken into account}'' by the
present authors. 
We attempt to clarify the relevant issues
pertaining to our analytical expressions, in order to indicate explicitly
the approximations made in the derivations and resolve problems with their
interpretation. Importantly, we also demonstrate that the analytical results
for amplitudes of combination harmonics are numerically accurate even in
most unfavorable circumstances at zero temperature and for clean samples.

The basic equations of the theory \cite{alebra3} are those for the
oscillating part $\tilde{F}$ of the free energy, which is the thermodynamic
potential of the canonical ensemble, 
\begin{equation}
\tilde{F}=\tilde{\Omega}(\mu )-(2\rho )^{-1}(\partial \tilde{\Omega}%
/\partial \mu )^{2},  \label{eq:Ft}
\end{equation}
and for the chemical potential $\mu =\mu _{0}+\tilde{\mu}$ , where $\tilde{%
\mu}=\rho ^{-1}\partial \tilde{\Omega}/\partial \mu $ [Eq.~(12) and Eq.~(9)
of Ref. \cite{alebra3}, respectively]. Here $\mu _{0}=\rho
^{-1}(N+\sum_{\alpha }\rho _{\alpha }\Delta _{\alpha })$ is the zero-field
chemical potential, $\rho $ and $\rho _{\alpha }$ are total and partial
zero-field densities of states (DOS), and $\Delta _{\alpha }$ is the $\alpha
-$band edge. The oscillating part of the grand canonical potential $\tilde{%
\Omega}$ (Eq.~(6) of \cite{alebra3}) is given by the standard expression 
\cite{sho} 
\begin{equation}
\tilde{\Omega}(\mu )=2\sum_{\alpha }\sum_{r=1}^{\infty }(-1)^{r}A_{\alpha
}^{r}\cos \left( \frac{rf_{\alpha }}{B}\right) ,
\end{equation}
where $A_{\alpha }^{r}$ are the amplitudes of the single-band Fourier
harmonics. For the sake of simplicity, we take the spin-splitting $g-$%
factors to be zero. Substituting this expression into $\tilde{F}$, one
obtains the combination amplitudes $C_{\alpha \alpha ^{\prime }}^{rr^{\prime
}}$ as [5] 
\begin{equation}
C_{\alpha \alpha ^{\prime }}^{rr^{\prime }}=2\pi ^{2}\frac{rr^{\prime
}A_{\alpha }^{r}A_{\alpha ^{\prime }}^{r^{\prime }}}{\rho \omega _{\alpha
}\omega _{\alpha ^{\prime }}},  \label{eq:C}
\end{equation}
where $\omega _{\alpha }=eB/m_{_{\alpha }}.$ Our definition of the
frequencies $f_{\alpha }=2\pi m_{_{\alpha }}(\mu -\Delta _{\alpha })/e$ in
Eq.~(2) and in Ref.~\cite{alebra3} contains the {\em exact} chemical
potential $\mu $ rather than its zero-field value $\mu _{0}.$ Hence,
Eqs.~(12) and (13) of Ref. \cite{alebra3} are {\em exact}$.$ They fully take
into account the chemical potential oscillations in the arguments of the
Fourier components.

Certainly, we did not consider our explicit expression for $\tilde{F}$,
Eqs.~(12) and (13)\cite{alebra3} as a final Fourier series, as should be
obvious from our using the exact $\mu ,$\ not its smooth part $\mu _{0},$\
in the expression for the free energy. The free energy $\tilde{F}$ has been
expanded in powers of $\tilde{\mu}\ll \mu _{0}$ as 
\begin{equation}
\tilde{F}\approx \tilde{\Omega}(\mu _{0})+(2\rho )^{-1}(\partial \tilde{%
\Omega}(\mu _{0})/\partial \mu _{0})^{2},
\end{equation}
and, differentiating it with respect to magnetic field, we have obtained the
result, Eq.~(16) \cite{alebra3}, for the magnetization amplitudes. The ratio
of the combination $M_{\alpha \alpha ^{\prime }}^{11}$ and single band
(conventional) $M_{\alpha }^{1}$ \ Fourier amplitudes in a two-band metal
was found at $T=0$ to be (Eq.~(16) in \cite{alebra3}) 
\begin{equation}
\frac{M_{\alpha \alpha ^{\prime }}^{11}}{M_{\alpha }^{1}}=\frac{m_{\alpha }}{%
m_{\alpha }+m_{\alpha ^{\prime }}}\frac{f_{\alpha }\pm f_{\alpha ^{\prime }}%
}{f_{\alpha }}.
\end{equation}
It was generalized in Eqs.~(21), (22) of Ref. \cite{alebra4} by taking into
account a background DOS, the Dingle, spin, and Yamaji reduction factors.
Obviously, the same result for the magnitude of this ratio can be obtained
by differentiating our exact free energy, Eqs.~(12,13) in \cite{alebra3},
with respect to the magnetic field $B$ but keeping frequency $f_{\alpha }$
constant. Therefore, the coefficients $C_{\alpha \alpha ^{\prime
}}^{rr^{\prime }}$ in the second term of the free energy indeed yield the
correct combination frequency amplitudes in the magnetization, as we pointed
in our original paper \cite{alebra3}. The expansion of the free energy in
powers of $\tilde{\mu}$ described above is straightforward, and it was not
mentioned explicitly in \cite{alebra3}. The present discussion should
clarify the point that the combination frequencies are fully defined by the
second-order expansion coefficients $C_{\alpha \alpha ^{\prime
}}^{rr^{\prime }}$.

Alternatively one can first differentiate the free energy, Eq.~(1) with
respect to $B$ as\cite{alebra}, 
\begin{equation}
\frac{\partial \tilde{F}}{\partial B}=\left( \frac{\partial \tilde{\Omega}}{%
\partial B}\right) _{\mu }+\frac{\partial \tilde{\Omega}}{\partial \mu }%
\frac{\partial \tilde{\mu}}{\partial B}-\frac{1}{\rho }\frac{\partial \tilde{%
\Omega}}{\partial \mu }\frac{\partial }{\partial B}\frac{\partial \tilde{%
\Omega}}{\partial \mu }=\left( \frac{\partial \tilde{\Omega}}{\partial B}%
\right) _{\mu },
\end{equation}
which gives the exact expression for the magnetization (Eq.~(5) of our
original paper \cite{alebra}). Then one can expand the result in powers of $%
\tilde{\mu}$, as it was done by Champel in the second part of his paper\cite
{cham}. Because the derivatives with respect to $\mu $ and with respect to $B$
commute, the approximate amplitudes should be the same as in Eq.~(5).
Indeed, the exact derivation of the magnetization, the approximate Fourier
amplitudes, and the ``main'' result Eq.~(14) of Ref. \cite{cham} are {\em %
identical} to our original expressions, Eq.~(5) of Ref.\cite{alebra} and
Eq.~(16) \cite{alebra3}, Eqs.~(21) and (22) \cite{alebra4}, respectively.

Champel mentions that ``...{\em the mechanism responsible for the
combination frequencies in the Fourier spectrum of magnetization
oscillations cannot apriori depend on the way the magnetization is derived,
that is to say, on the use of a specific thermodynamic potential. Our
following goal is then to point out how the combination frequencies arise by
considering directly the expression for the magnetization oscillations in
the relevant thermodynamic limit}''. Indeed, our results do not
depend on the use of a particular thermodynamic potential. The mechanism of
novel combination frequencies\cite{alebra} is based on the effect of
chemical potential oscillations, which is important in quasi-2D closed
metallic systems and is absent in open systems. It can be derived with the
use of either thermodynamic potential, its particular selection being a
matter of convenience. To imply otherwise would be to misrepresent our work.
Obviously, a natural choice of a thermodynamic potential for an open system
(with a constant chemical potential) is $\Omega $($\mu $), while for a
closed system with the constant number of particles $N$\ this would be $F(N)$%
. One is free to use $\Omega (\mu )$\ at constant $N$\ as far as one
accounts for the functional relation $\mu =\mu (N)$\ in the derivation,
although it only makes the derivation cumbersome without changing any
results. The author of \cite{cham} seems to realize this, in fact using our
formulas, e.g. Eqs.~(3)\ and (4) in \cite{cham}, to rederive our prior
results.like e.g. (14) in Ref.\cite{cham}.

The difference between the corresponding free energies $F$and $\Omega
$ of
the measured system is tiny, since it is proportional to the fluctuation of
the carrier density, Eq.~(\ref{eq:Ft}). However, the effect on
susceptibility is greatly amplified, since two differentiations with respect
to the field bring about very large factor $\left( F/B\right) ^{2}\gg 1$,
Ref.\cite{alebra4}, similar to the case of the usual de Haas-van Alphen
effect. Obviously, there will be no chemical potential oscillations when it
is fixed by a reservoir, so that no combination frequencies due to these
oscillations can be observed in an open (grand canonical) system. The
reservoir commonly implies a large system with an infinite continuum of
nonquantized electron states. In this sense the role of the reservoir may be
played by e.g. a one-dimensional (and, therefore, not closed) electron orbit
with very large ``background'' density of states in the same quasi-2D sample 
\cite{alebra2}, observed in some cases. On the contrary, the unusual
combination frequencies appear when the system is closed, so that the number
of carriers stays the same, and the chemical potential must oscillate \cite
{alebra}. In actual dHvA experiments the sample is usually measured while
placed on non-conducting substrate with no electrodes attached, so the
system is indeed closed. Note that the definition of the chemical potential
does {\em not} require that the system be open. For normal Fermi liquids
with the ground state energy $E_{0}(k)$ for system of $k$ particles the
chemical potential $\mu =E_{0}(k+1)-E_{0}(k)$ is uniquely defined for large $%
k $.

Further, Champel writes about the origin of combination frequencies: ``...%
{\em \ in 2D metals,\ the magnetization oscillations become more or less
sensitive to the presence of the chemical-potential\ oscillations and
exhibit significantly different behaviors depending on the presence or
absence of a finite reservoir\ of electrons.[2,9,10] \ In 2D multiband
metals this higher sensitivity to chemical-potential oscillations is
expressed by the presence of the combination frequencies in the Fourier
spectrum of magnetization\ oscillations at very low temperatures, as shown
numerically by Nakano [11]. Here, our aim is to prove analytically the
existence of these combination frequencies in the magnetization oscillations}%
''. One would note that the first claim is well known for single-band 2D
metals since the pioneering work by Peierls in 1933
\cite{peierls}. One would easily see 
that the existence of combination frequencies has been proven numerically in
our work \cite{alebra} prior to Nakano's work. The analytical proof of their
existence has already been given in Refs.\cite{alebra3,alebra4}.

\begin{figure}[t]
\epsfxsize=3.4in \epsffile{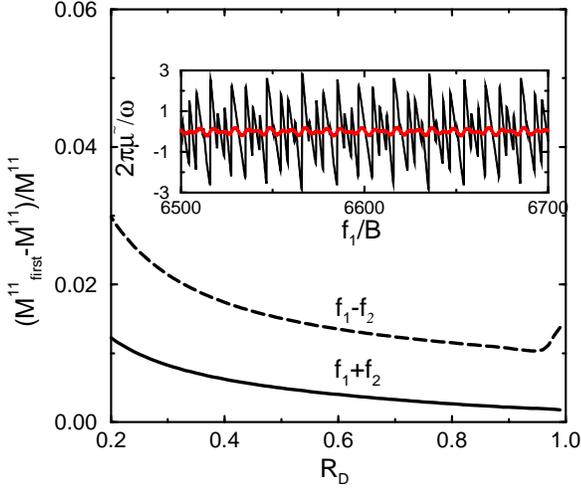}
\caption{ Relative ratio of the first approximation $M_{first}^{11}$, Eq.~(%
\ref{eq:Mfirst}), and exact $M^{11}$ combination amplitudes for a range of
Dingle factors $R_{D}$. Note that the difference is in the range of 1
percent or less for both dirty and clean systems. We have assumed $%
f_{2}/f_{1}=7/11$. Inset: the oscillating part of the chemical potential $z=2%
\protect\pi \tilde{\protect\mu}/\protect\omega $ as a function of inverse
magnetic field $1/B$. The two curves correspond to clean (the Dingle factor $%
R_{D}=0.99$, large amplitude of chemical potential oscillations $2\protect\pi
\tilde{\protect\mu}/\protect\omega $) and dirty ($R_{D}=0.1$, small
amplitude) samples, respectively. }
\label{fig:fig1}
\end{figure}

Now we would like to address the claim of Ref. \cite{cham} that ``{\em the
quantitative description of dHvA oscillations in terms of a Fourier series
may break down and has to be done numerically}''. Champel \cite{cham} has
failed to mention that the numerical description of combination harmonics
was done in our original \cite{alebra} and subsequent papers \cite
{alebra2,alebra4}. To illustrate the accuracy of the analytical theory
[5,7], let us analyze a simple two-band model. The results below show
explicitly that the higher powers of $\tilde{\mu}$ -expansion can be
neglected even at zero temperature and in a clean sample. Consider two
bands with equal masses $m_{1}=m_{2}\equiv m$ and equal Dingle reduction
factors $R_{D}=\exp (-2\pi \Gamma /\omega ),$ but with the different dHvA
frequencies $f_{1}$ and $f_{2}.$ This two-band model is the most unfavorable
case for the first-order expansion because in the presence of more bands the
oscillations of the chemical potential are reduced compared to the two-band
case, as was correctly noted by Champel \cite{cham}. The derivative of $%
\tilde{\Omega}$, Eq.~(2) with respect to $\mu $ and the derivative of $%
\tilde{F},$ Eq.~(6) with respect to $B$ yield for $T=0$ 
\begin{equation}
z=-\sum_{r=1}^{\infty }\frac{(-R_{D})^{r}}{r}\left[ \sin (rz+rf_{1}/B)+\sin
(rz+rf_{2}/B)\right] ,
\end{equation}
and 
\begin{eqnarray}
\tilde{M} &=&\frac{e^{2}}{4\pi ^{3}m}\sum_{r=1}^{\infty }\frac{(-R_{D})^{r}}{%
r} \\
&&\times \left[ f_{1}\sin (rz+rf_{1}/B)+f_{2}\sin (rz+rf_{2}/B)\right] , 
\nonumber
\end{eqnarray}
where $z\equiv 2\pi \tilde{\mu}/\omega ,$ and $\tilde{M}$ is the oscillating
part of the magnetization ($\omega =eB/m$). Expanding $\tilde{M}$ in powers
of $z$ and neglecting $z$ in the right hand side of Eq.~(7) yields the
amplitudes of the ($f_{1}\pm f_{2})$ Fourier components as \cite{alebra3} 
\begin{equation}
M_{first}^{11}=\mp \frac{e^{2}R_{D}^{2}}{16\pi ^{3}m}(f_{1}\pm f_{2}).
\label{eq:Mfirst}
\end{equation}
One can estimate the accuracy of this expression by Fourier transforming
Eq.~(8) without the expansion but leaving only the leading $(r=1)$ harmonics
in the chemical potential, $z\approx R_{D}\left[ \sin (f_{1}/B)+\sin
(f_{2}/B)\right] $. Then the ratio of the ``exact'' combination amplitude to
the approximate first-order amplitude is given by 
\begin{equation}
\frac{M^{11}}{M_{first}^{11}}\approx \frac{2J_{0}(R_{D})J_{1}(R_{D})}{R_{D}},
\end{equation}
where $J_{n}(x)$ is the Bessel function. This ratio is very close to unity
at any value of the Dingle factor. In fact, the first order expansion \cite
{alebra3} has even better accuracy than the estimate given by Eq.~(10).

For the numerical calculations we have used the analytical expressions of
the sums in Eqs.~(7) and (8), 
\begin{equation}
z={\rm Arg}\left\{ \left[ 1+R_{D}e^{i(z+f_{1}/B)}\right] \left[
1+R_{D}e^{i(z+f_{2}/B)}\right] \right\} ,
\end{equation}
and 
\begin{eqnarray}
\tilde{M} &=&-\frac{e^{2}f_{1}}{4\pi ^{3}m}  \nonumber \\
&&\times {\rm Arg}\left\{ \left[ 1+R_{D}e^{i(z+f_{1}/B)}\right] \left[
1+R_{D}e^{i(z+f_{2}/B)}\right] ^{f_{2}/f_{1}}\right\} ,
\end{eqnarray}
where {\rm Arg}$w$ stands for the argument of the complex number $w.$ The
oscillating part of the chemical potential and the relative difference
between the analytical and numerical amplitudes are shown in Fig.~\ref
{fig:fig1}. The first-order Fourier amplitudes of the magnetization are
equal to the numerical amplitudes within a few percent at any value of the
collision broadening $\Gamma .$ According to our theory, higher {\em %
combination} harmonics, like $f_{1}+2f_{2}$, should be exponentially
suppressed compared to the leading harmonics $f_{1}\pm f_{2}$ studied here.
It is not surprising, therefore, that the higher combination harmonics have
not been clearly resolved even in very clean samples\cite{ohm99}.{\rm \ }%
Evidently, the present results for the combination $f_{1}\pm f_{2}$\
harmonics are sufficient to refute Champel's general claim that our
analytical theory looses accuracy at low temperatures in clean systems.
Indeed, the leading analytical combination amplitudes \cite{alebra3,alebra4}
are very accurate even at low temperatures and in clean samples, contrary to
claims in Ref. \cite{cham}.

In conclusion, we have confirmed the validity of our analytical derivation 
\cite{alebra3}, which {\em does}{\it \ }take into account the chemical
potential oscillations. We have also shown that the experimentally observed
analytical combination amplitudes \cite{alebra3} of the magnetization are
numerically accurate even at zero temperature and even in clean samples.{\bf %
\ }Therefore, the present analysis justifies the use of the analytical
theory \cite{alebra3,alebra4} in the relevant range of all parameters.

\end{document}